\documentclass[a4paper,12pt]{article}
\usepackage{graphicx}
\usepackage{amssymb,latexsym}
\usepackage[english]{babel}
\usepackage{graphicx}
\newcommand{\bec}{\begin{center}}
\newcommand{\ec}{\end{center}}
\newcommand{\bee}{\begin{equation}}
\newcommand{\ee}{\end{equation}}

\begin{document}
\large
\begin{titlepage}
\begin{center}  
{\Large\bf {Application of the triangulated category to the explanation of fully-charm tetraquark mass}}\\
\vspace*{12mm} {\bf  T.V. Obikhod \\}
\vspace*{10mm}
{\it Institute for Nuclear Research\\
National Academy of
Sciences of Ukraine\\
03068 Kiev, Ukraine\\}
e-mail: obikhod@kinr.kiev.ua\\
\vspace*{21mm}
{\bf Abstract\\}
\end{center}
The discovery in July 2020 of fully-charm tetraquark led to the need for its theoretical explanation. For investigation of such complex four-quark formation, the modern mathematical apparatus of the theory of derived categories is used. By representing diquarks as solitonic objects in terms of sheaves, one can explain the measured mass of the broad resonance of fully-charm tetraquarks consisting of di-charmonia.
\vspace*{8mm}\\

\end{titlepage}

\begin{center} 
\textbf{\textsc{1. Introduction}}
\end{center}

	The searches for new physics beyond the standard model (SM) describing the interaction of all known elementary particles are the priority for the experiment at the Large Hadron Collider (LHC). In July 2020, a new exotic particle was discovered at the LHC (LHCb) facility - a tetraquark with a mass range of $6.2-7.4$ GeV, \cite{1.}. The observation was made with a confidence level of more than 5$\sigma$ (five standard deviations), which allows scientists to talk about the existence of a new, previously unknown particle.

	LHCb Collaboration reported their results on possible fully-charm tetraquarks as they investigated the di-J$/ \psi$ invariant mass spectrum, where they observed a broad structure ranging from 6.2 to 6.8 GeV and a narrow structure at around 6.9 GeV with a global significance of more than 5$\sigma$. The schematic representation of the formation of such a particle is shown in Fig. 1
\bec
\includegraphics[width=1\textwidth]{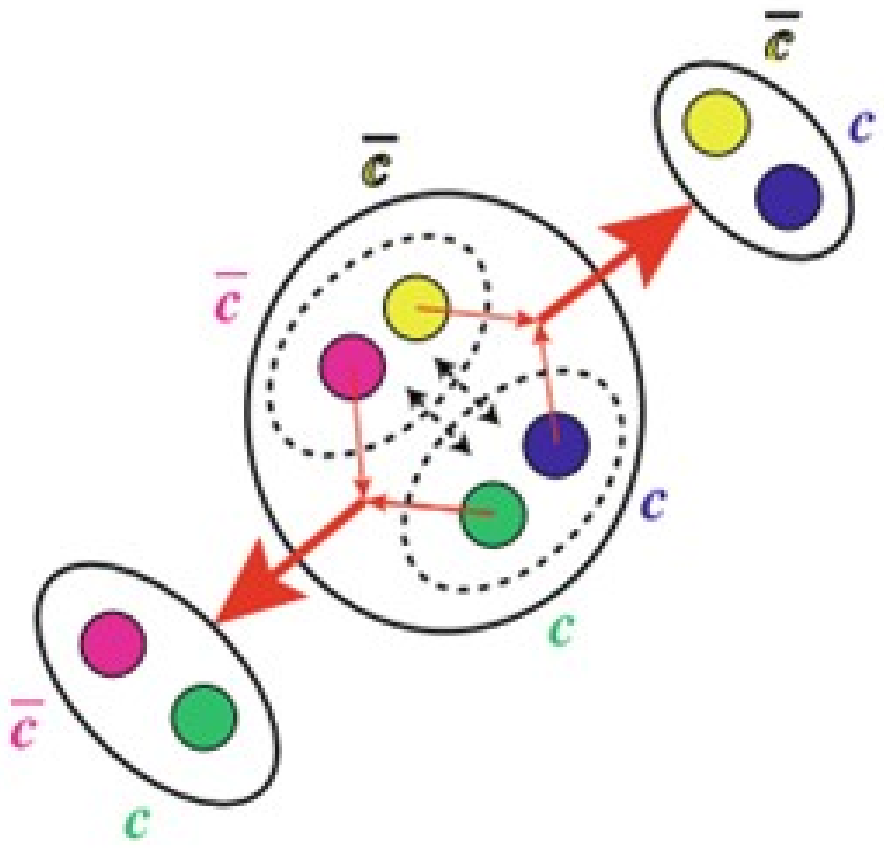}\\
\vspace{5 mm}
\emph{\textbf{Fig.1.}} {\emph{ The decay of a compact diquark-antidiquark state into two charmonium mesons. }}
\ec
The broad structure observed by LHCb at around $6.2–6.8$ GeV \cite{1.} can be interpreted
as an S-wave c$\overline{c}$c$\overline{c}$ tetraquark state. Accordingly, the narrow
structure observed by LHCb at around 6.9 GeV \cite{1.} can be interpreted as a P-wave 
c$\overline{c}$c$\overline{c}$ tetraquark state. "Bump" corresponds to a resonance 
(short-lived particle), which, according to forecasts, corresponds to a particle 
consisting of two charmed quarks and two charmed antiquarks, and therefore belongs to tetraquarks. 
The mass spectra of these two resonances is presented in Fig.2
\vspace{1 mm}
\bec
\includegraphics[width=0.8\textwidth]{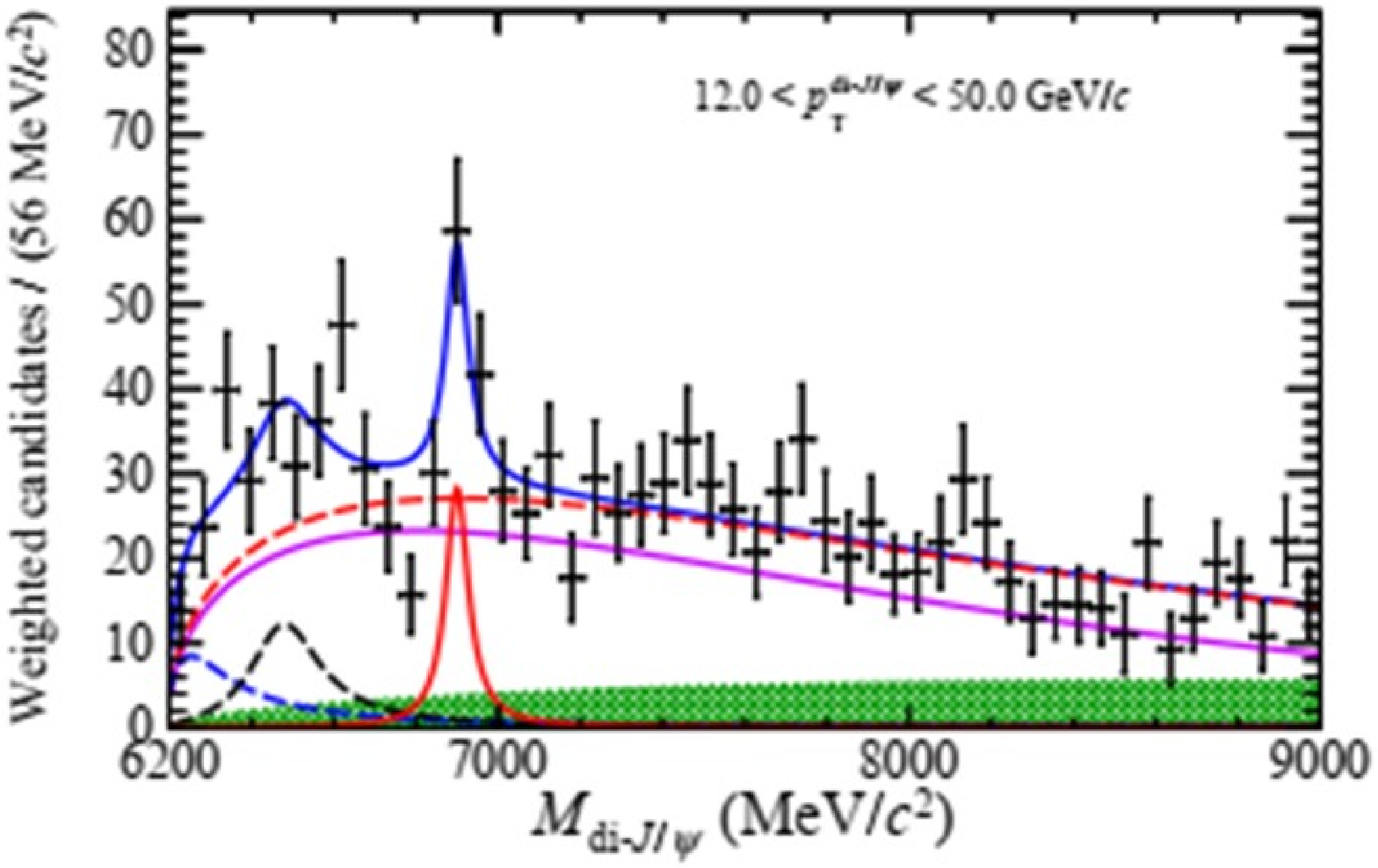}\\
\vspace{5 mm}
\emph{\textbf{Fig.2.}} {\emph{ Invariant mass spectra of weighted di-J$/ \psi$ candidates. }}\\
\ec
\vspace*{1mm}
	It is not yet clear whether the new particle is a real tetraquark, that is, a system of 
particles closely related to each other, or a "hadronic molecule" - two closely interacting mesons, 
consisting of a quark-antiquark pair. Predictions for the masses of 
c$\overline{c}$c$\overline{c}$ states vary from 5.8 to 7.4 GeV/c$^2$ \cite{2.,3.}, which are not in 
correspondence with the masses of known charmonia exotic states. This mass range is connected with the 
identification of discovered particle as J$/ \psi$ –pair. The mechanism for understanding the formation 
of such a state may be connected with two separate interactions of gluons or quarks, named double-parton 
scattering (DPS) or with single interaction, named nonresonant single-parton scattering (NRSPS). 
More data with aditional measurements, including determination of the spin-parity quantum numbers 
and p$_T$ dependence of the production cross-section, are needed to provide further information about 
the nature of the observed structure.

\begin{center} 
\textbf{\textsc{2. The mass of di-J$/ \psi$ tetraquark calculation}}
\end{center}

	The main mechanisms of formation of tetraquark states, provided by the diquark model and the molecular scheme, are considered in \cite{4.}. The properties of tetraquark states at large Nc are analyzed through the Feynman diagrams that describe two-meson scattering amplitudes. It turns out that the possible formation of tetraquark states is mainly due to the mutual interactions of their internal mesonic clusters. The spectrum of mesons is very similar to that of free strings, whose coupling constant would be of the order of $1/N_c$. As was noticed by tHooft, there exists the resemblance between QCD theory and string theory in the large-Nc limit. Therefore, for the of mesons, whose coupling constant is of the order of $1/N_c$, exists the duality relation between the string picture and QCD. Flat-space string theories are formulated in ten dimensions and they involve the scales of the order of the Planck scale, \cite{5., 6., 7.}. So, the clarification of the mechanisms that are at work in the formation of tetraquarks is the main from the large-N$_c$ approach.

	String theory has been present in hadronic physics from the early days of the discoveries of hadron resonances. Later, the advent of the quark model and of QCD confinement of quarks and gluons, became the reason for the introduction of strings, at the endpoints of which are attached quarks. In the aspect of AdS/CFT correspondence, N = 4 super-Yang-Mills (SYM) theories, in four dimensions, are found to be dual to type-IIB string theory in Anti-de Sitter (AdS) space, in five dimensions. The gauge theory corresponds to the compactified theory through a six-dimensional manifold M$^6$, while the string theory is compactified through a five-dimensional sphere S$^5$, leading to the correspondence of spaces: 
\[	R^{1,3}\times M^6 \Leftarrow \Rightarrow AdS^5\times S^5. \]

	The most important development in string theory is the discovery of D-brane as a subspace of the target spacetime on which open strings may end \cite{8.}. So, it would be interesting to find the elementary particles, which are treated as vibrational modes of strings located between D-branes and connected with a consistent quantum theory. It was proposed by Aspinwall \cite{8.} the application of the theory of derived category for the definition of stability properties of such objects. The objects of such category are D-branes and morphisms are strings, connecting these D-branes.

	In the general case, the D-branes correspond to the complex
\[\ldots \rightarrow E^{-1}\stackrel{d_{-1}}{\rightarrow}E^0\stackrel{d_0}{\rightarrow}E^1 \stackrel{d_1}{\rightarrow}\ldots\ ,\]
where $E^n$ - coherent sheaves, $d_n$ - morphisms between sheaves, that satisfy the 
condition $d^2 = 0$. The set of complexes and morphisms between them form a derived
 category D(X) of coherent sheaves on X - the space of extra dimensions or M$^6$.  
Two D-branes are compexes $E^{\bullet}, F^{\bullet} $
in D(X) connected by morphism $f: E^{\bullet}\rightarrow F^{\bullet}$  in
Cone($f: E^{\bullet}\rightarrow F^{\bullet}$). This leads us to study the structure 
of the category of topological D-branes as a triangulated category. 
Triangle in D(X) - sequence of morphisms connecting vertices $\bf{A}$, $\bf{B}$ and $\bf{C}$ as shown below  
\vspace{1 mm}
\bec
\includegraphics[width=0.5\textwidth]{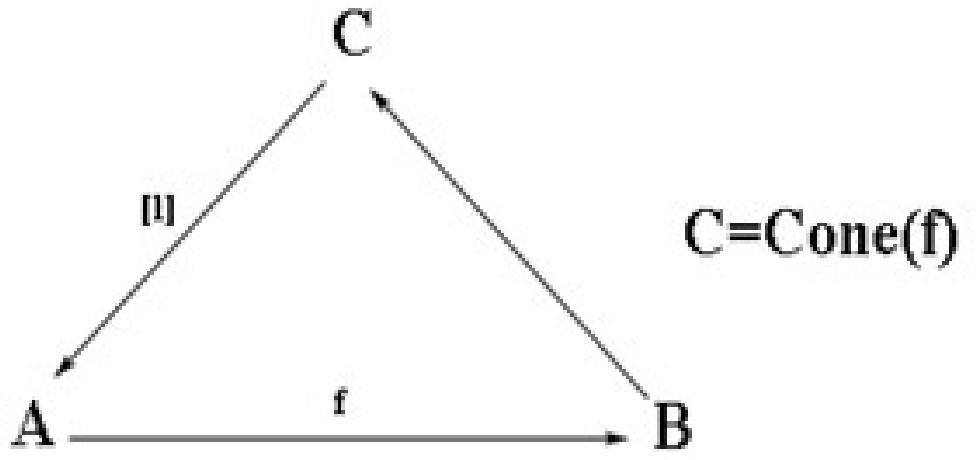}\\
\vspace{5 mm}
\ec
\vspace*{1mm}
	We turn to the problem of phase transition or stability of D-branes, which are potential bound states of fractional D-branes. Let's use the construction of distinguished triangles, \cite{9.}. To determine the stability of the sheaf $\bf{C}$, we use the grading concept where Z($\bf{A}$) is central charge and $\xi(\bf{C})$ is grading of sheaf $\bf{A}$
\[\xi(A)=\frac{1}{\pi}arg Z(A)\]
\[\xi(B)=\frac{1}{\pi}arg Z(B)\]
To study the stability of the sheaf $\bf{C}$ it is necessary to calculate the square of the mass of the string $\bf{f}$ connecting sheaves $\bf{A}$ and $\bf{B}$, \cite{10.}:
\[m^2=\frac{1}{2}(\xi(B)-\xi(A)+q-1)\]
The criterion of stability of the D-brane $\bf{C}$ says:\\
$\bullet$ D-brane is stable with respect to decay into D-branes $\bf{A}$ and $\bf{B}$ if $q<1$;\\
$\bullet$ D-brane is unstable with respect to decay into D-branes $\bf{A}$ and $\bf{B}$ if $q>1$.\\
Suppose we have the distinguished triangle with $\bf{A}$ and $\bf{B}$ – J$/ \psi$ particles. We can consider bound state Cone(f) of these two particles. As tetraquark is broad structure at around $6.2–6.8$ GeV interpreted as an S-wave c$\overline{c}$c$\overline{c}$ tetraquark state, we can say, that $q<1$ for the mass of string connecting two sheaves $\bf{A}$ and $\bf{B}$
\vspace{1 mm}
\bec
\includegraphics[width=0.8\textwidth]{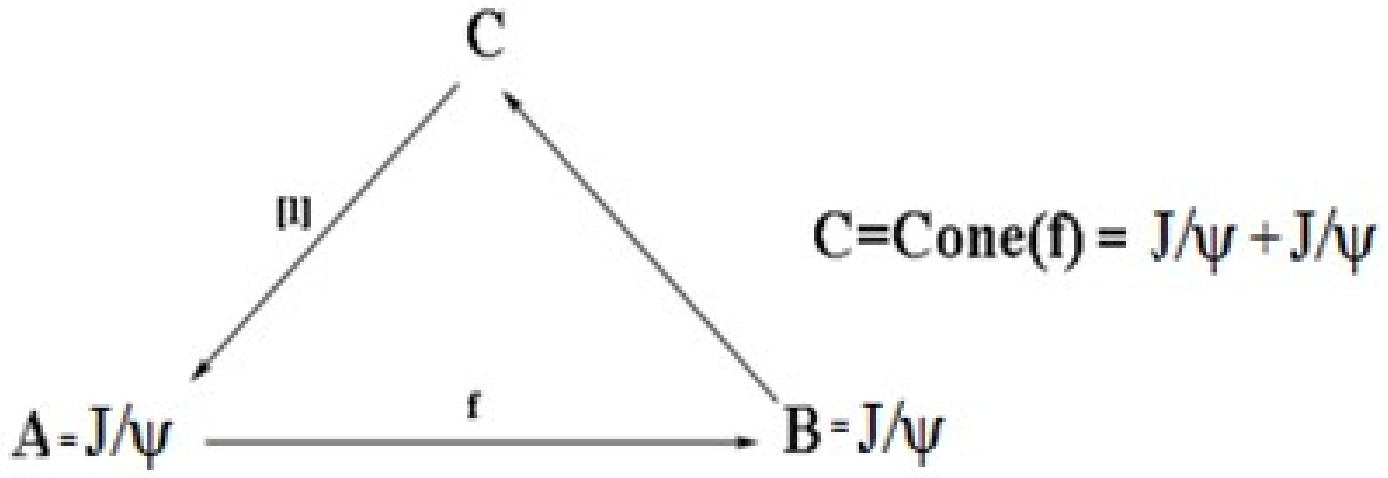}\\
\vspace{5 mm}
\ec
\vspace*{1mm}
The mass of such object is equal to the sum of masses of J$/ \psi$ particles with masses of about 3.1 GeV and equal to 6.2 GeV. These data are in agreement with observed mass range of broad tetraquark, $6.2-6.8$ GeV, obtained experimentally.

\begin{center} 
\textbf{\textsc{3. Conclusions}}
\end{center}
The discovery of tetraquarks and, in particular, di-J$/\psi$ is a striking experimental event at the LHC. Its existence does not contradict the Standard Model, but the explanation of its mass within the framework of the Lagrangian formalism is a difficult task associated with solving complex nonlinear equations. Therefore, it is relevant to calculate the masses of such objects using the theory of derived categories - a powerful mathematical apparatus of high energy physics. Among the theories of derived categories, the theory of category of distinguished triangles is widely used. As part of our work, we considered a selected triangle with vertices - J/$\psi$ particles and obtained a mass consistent with the mass of an experimentally discovered particle. The  obtained result confirms the need to use the mathematical apparatus of high energy physics - the theory of superstrings and D-branes to explain modern experimental discoveries.

\end{document}